\documentclass[preprint2]{aastex}
\slugcomment{Submitted to astro-ph.}
\usepackage{wasysym}
\def\crossarrow{%
\mathop{\vcenter{\hbox{\ooalign{\raise -0.1ex%
        \hbox{$\hbox{$\longrightarrow$\hspace*{.05em}}$}\cr\cr%
                $\hbox{\hspace*{.6em}/}$}}}}}%

\begin{document}

%\title{Chemistry in the inner regions of protoplanetary disks\\
%1. The fractionation of carbon isotopologues (carbon-bearing species?)}
\title{Carbon isotope measurements in the Solar System}

\author{Paul M. Woods} \affil{ Jodrell Bank Centre for Astrophysics,
  Alan Turing Building, School of Physics \& Astronomy\\ The
  University of Manchester, Oxford Road, Manchester M13 9PL, UK}
\email{paul.woods@manchester.ac.uk}

\begin{abstract}
I make publicly available my literature study into carbon isotope
ratios in the Solar System, which formed a part of \citet{woo09}. As
far as I know, I have included here all measurements of
$^{12}$C/$^{13}$C in Solar System objects (excluding those of Earth)
up to and including 1 February 2010. Full references are given. If you
use the any of the information here, please reference the paper 
\citet{woo09} and this publication.
\end{abstract}


\section{Introduction}

\begin{figure*}
\epsscale{2.0}
\plotone{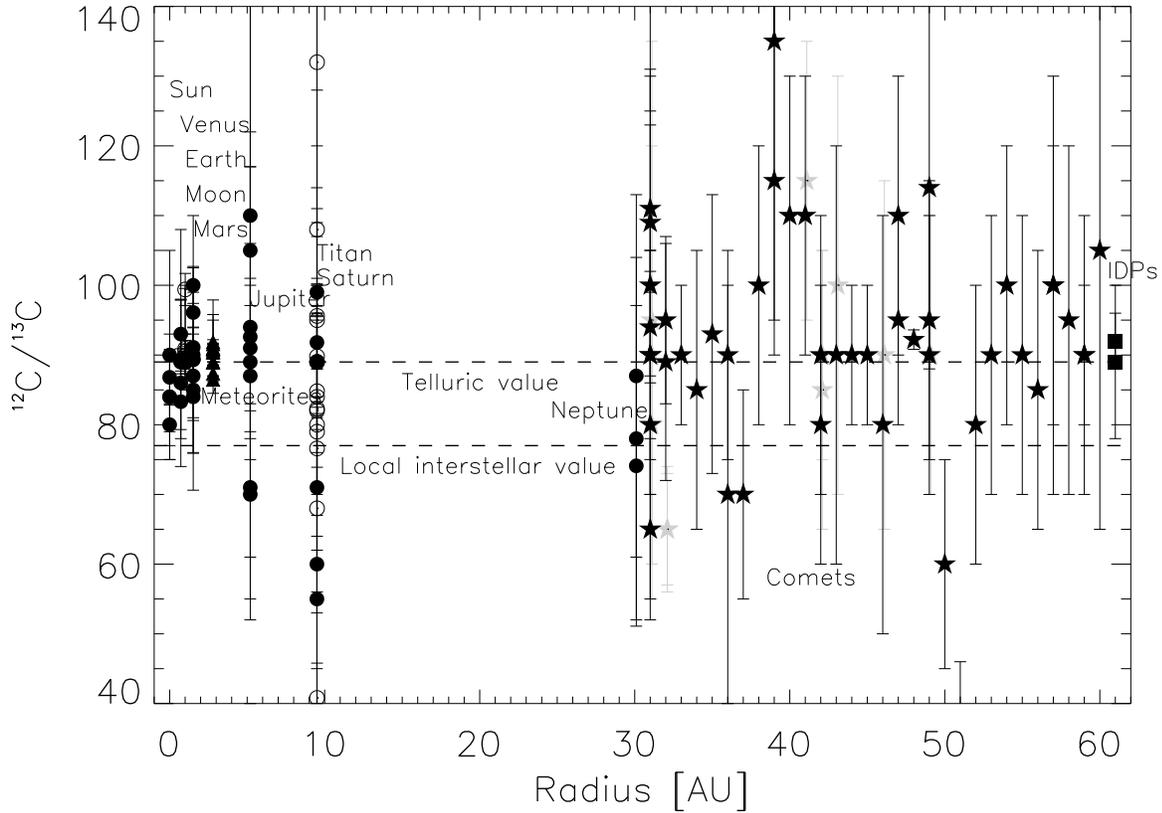}
\caption{Measurements of the $^{12}$C/$^{13}$C ratio in various
  objects of the Solar System. Filled circles indicate measurements of
  planets or the Sun and empty circles indicate measurements of
  planetary moons. Triangles indicate bulk isotope measurements of the
  $^{12}$C/$^{13}$C ratio in meteorites, and have been placed at the
  radius of the asteroid belt. Comets, indicated by filled stars, have
  been placed outside of the radius of Neptune, for
  illustration. Those cometary points in light grey show depreciated
  measurements, which have been superseded by updated calculations
  (this is not a conclusive list, and stems mainly from
  \citet{man09}). IDPs (filled squares) have been placed at cometary
  radii to indicate their likely origin in comets.}
\label{fig:SSiso}
\end{figure*}

In the course of writing \citet{woo09}, I read a large number of
papers on carbon isotopes in the Solar System, and compiled as much
data as I could about these isotope ratios, $^{12}$C/$^{13}$C , into a
figure: Fig.~10 of that paper. An updated version of this figure can
be found as Fig.~\ref{fig:SSiso} in this paper. Here I list all the
reference sources I was able to find in the hope that it will help
other parties interested in carbon isotope ratios.

Ratios often have to be converted from the delta notation, favoured by
meteoriticists. This is simply,
\begin{equation}
\delta {^{13}\mathrm{C}} =
\left(\frac{{^\mathrm{13C}R_\mathrm{meas}}}{{^\mathrm{13C}R_\mathrm{std}}} -1\right)
  \times 1000,
\end{equation}
expressed in permil ($\permil$). ${^\mathrm{13C}R_\mathrm{meas}}$ is
the measured value of the $^{13}$C/$^{12}$C ratio,
${^\mathrm{13C}R_\mathrm{std}}$ is the terrestrial standard, often
taken to be the Peedee belemnite (PDB) value of 0.0112372. This
quantity can often appear inverted, and particularly in astrophysical
literature can be quoted as $^{12}$C/$^{13}$C$_\mathrm{std}$=89. I use
this convention here.  Uncertainties in reported values are sometimes
given in the literature, and I try to include those where
possible. Also, ranges in values may be stated when a number of
similar measurements have been made. In the tables of data,
Tables~\ref{tab:innerplanets}-\ref{tab:comets}, I have interpreted
these ranges as errors about the median point. References are given in
order that the ratios given in the tables may be verified - I do not
claim that these results are free from mistakes. Neither do I claim
that this list of data or references is complete. I believe it to be a
fairly complete sample of what is available in the astrophysical
literature up until February 2010.

There are four tables of data in this paper:
Table~\ref{tab:innerplanets} gives $^{12}$C/$^{13}$C ratios for the
Sun, moon and rocky planets. Table~\ref{tab:outerplanets} provides
data on the outer planets, including the molecule(s) observed to make
the $^{12}$C/$^{13}$C ratio
determinations. Table~\ref{tab:minorbodies} contains $^{12}$C/$^{13}$C
ratios for minor Solar System bodies, such as meteorites and their
incorporated pre-solar grains, and interplanetary dust particles
(IDPs). The final table, Table~\ref{tab:comets}, gives carbon isotope
ratios in comets. Some of the ratios in this table are a result of
re-examining the same observational data with new techniques or
models. For example, the $^{12}$C/$^{13}$C ratio in comet Halley was
determined to be 65 by \citet{wyc89}; however, the data was
re-evaluated with a new model by \citet{kle95}, resulting in
$^{12}$C/$^{13}$C=95. For completeness, I keep both results in the
table. \citet{sad96} contains a useful discussion of revisions to
estimates of $^{12}$C/$^{13}$C in the outer
planets. Section~\ref{sec:context} gives some context into the matter
of carbon isotope ratios, and then data tables follow.

\section{Carbon isotope data}

\subsection{Interstellar and Solar System context}
\label{sec:context}

The isotope ratio for carbon ($^{12}$C/$^{13}$C) in the Solar System
is widely accepted to be 89 \citep{and89,cla04,mei07}, although recent
measurements of the solar photosphere have indicated a ratio of
80$\pm$1 \citep{ayr06}. This is greater than in the local interstellar
medium (ISM), where the value is taken to be 77 \citep{wil94}, greater
than the Orion Bar region
\citep[$^{12}$C/$^{13}$C$\sim$60;][]{kee98,lan90}, and much greater
than the Galactic Centre
\citep[$^{12}$C/$^{13}$C$\sim$20;][]{mil05,lan84}. This galactic
gradient \citep{lan90} is due to the higher star formation rate in the
inner Galaxy \citep{tos82}, where the fraction of $^{13}$C has been
enhanced by the $^{13}$C-rich ejecta of evolved, intermediate-mass
stars \citep{ibe83} in the time since the formation of the Solar
System.

%\clearpage

\subsection{Planetary bodies and the Sun}

See Tables~\ref{tab:innerplanets} and \ref{tab:outerplanets} for data
on the minor and major planets, respectively.

\begin{deluxetable}{lccl}
\tablecaption{\label{tab:innerplanets} $^{12}$C/$^{13}$C ratios in
  the inner planets and the Sun.}  \tablehead{\colhead{Object} &
  \colhead{$^{12}$C/$^{13}$C} & \colhead{Error} & \colhead{Reference(s)}} 
\tablewidth{0pt}
\startdata
Sun & 80   & +3/-3   & \citet{ayr06} \\
    & 86.8 & +4/-4   & \citet{sco06a} \\
    & 84   & +5/-5   & \citet{har87} \\
    & 84   & +9/-9   & \citet{hal73} \\
    & 90   & +15/-15 & \citet{hal72} \\
Venus & 86   & +12/-12   & \citet{bez87} \\
      & 93   & +15/-15   & \citet{bal86} \\
      & 89.3 & +1.6/-1.6 & \citet{ist80} \\
      & 83.3 & +4/-4     & \citet{hof79} \\
      & 89   & --        & \citet{nie79} \\
      & 89   & --        & \citet{con68} \\
Earth$^\dag$ & 89.0 & +0.4/-0.4 & \citet{hol83} \\
      & 91.9 & +3.9/-3.9 & \citet{sco06b} \\
      & 89.4 & +0.2/-0.2 & \citet{cop87} \\
      & 89   & +4/-4     & \citet{wed69} \\
Moon & 99.4 & +2.3/-2.3 & \citet{has04} \\
     & 89.4 & +2.7/-2.7 & \citet{wie04} \\
     & 89.9 & +0.3/-0.3 & \citet{bec82} \\
     & 89.4 & +2.2/-2.1 & \citet{bec80} \\
     & 89   & +2.2/-2.3 & \citet{eps72} \\
     & 89.6 & +0.5/-0.5 & \citet{eps71} \\
     & 90.5 & --        & \citet{cha71} \\
     & 89.9 & +0.6/-0.6 & \citet{kap71} \\
     & 89   & +1.0/-1.0 & \citet{eps70} \\
     & 90.9 & --        & \citet{kap70} \\
%     &      &          & \citet{moo70} \\
Mars & 91.1 & +1.9/-1.8   & \citet{kra07} \\
     & 89.4 & +11.0/-8.8  & \citet{enc05} \\
     & 84.0 & +13.4/-13.4 & \citet{kra96} \\
     & 96.1 & +6.4/-5.7   & \citet{sch86} \\
     & 89.4 & +4.5/-4.5   & \citet{nie77} \\
     & 89.3 & +13.4/-13.4 & \citet{mag77} \\
     & 90   & +9/-9       & \citet{owe77} \\
     & 85   & +9/-9       & \citet{bie76} \\
     & 87   & +3/-3       & \citet{nie76} \\
     & 100  & --          & \citet{kap69}
\enddata
\tablecomments{$^\dag$ no attempt is made to make this a complete list of measurements for the Earth.}
\end{deluxetable}
\begin{deluxetable}{lcccl}
\tablecaption{\label{tab:outerplanets} $^{12}$C/$^{13}$C ratios in the
  outer planets.}  \tablehead{\colhead{Object} &
  \colhead{$^{12}$C/$^{13}$C} & \colhead{Error} & \colhead{Molecule} &
  \colhead{Reference(s)}} 
\tablewidth{0pt}
\tabletypesize{\small}
\startdata
Jupiter & 92.6 & +4.5/-4.1 & CH$_4$     & \citet{nie98,nie96} \\
        & 160  & +40/-55   & CH$_4$     & \citet{cou83} \\
        & 87   & +35/-35   & CH$_4$     & \citet{cou83} \\
        & 71   & +12/-10   & CH$_4$     & \citet{cou83} \\
        & 89   & +12/-10   & CH$_4$     & \citet{com79} \\
        & 89   & +12/-10   & CH$_4$     & \citet{com77} \\
        & 70   & +30/-15   & CH$_4$     & \citet{deb76} \\
        & 110  & +35/-35   & CH$_4$     & \citet{fox72} \\
        & 105  & +12/-12   & C$_2$H$_6$ & \citet{sad96} \\
        & 91   & +26/-13   & C$_2$H$_6$ & \citet{sad96} \\
        & 94   & +12/-12   & C$_2$H$_6$ & \citet{wie91} \\
        & 20   & +20/-10   & C$_2$H$_6$ & \citet{dro85} \\
Saturn  & 91.8 & +8.4/-7.8 & CH$_4$     & \citet{fle09} \\
        & 71   & +25/-18   & CH$_4$     & \citet{cou83} \\
        & 89   & +25/-18   & CH$_4$     & \citet{com77} \\
        & 55   & +40/-15   & CH$_4$     & \citet{lec76} \\
        & 60   & +40/-15   & CH$_4$     & \citet{com75} \\
        & 99   & +43/-23   & C$_2$H$_6$ & \citet{sad96} \\
Titan   & 84   & +17/-17   & CO$_2$     & \citet{nix08b} \\
        & 76.6 & +2.7/-2.7 & CH$_4$     & \citet{nix08a} \\
        & 82.3 & +1/-1     & CH$_4$     & \citet{nie05} \\
        & 95.6 & +0.1/-0.1 & CH$_4$     & \citet{wai05} \\
        & 82   & +27/-18   & CH$_3$D    & \citet{bez07} \\
        & 89   & +22/-18   & HCN        & \citet{vin07} \\
        & 68   & +16/-12   & HCN        & \citet{vin07} \\
        & 132  & +25/-25   & HCN        & \citet{gur04} \\
        & 108  & +20/-20   & HCN        & \citet{gur04} \\
        & 95   & +25/-25   & HCN        & \citet{hid97} \\
        & 79   & +17/-17   & HC$_3$N    & \citet{jen08} \\
        & 84.8 & +3.2/-3.2 & C$_2$H$_2$ & \citet{nix08a} \\
        & 89   & +8/-8     & C$_2$H$_6$ & \citet{jen09} \\
        & 89.8 & +7.3/-7.3 & C$_2$H$_6$ & \citet{nix08a} \\
        & 80   & +20/-20   & C$_2$H$_6$ & \citet{ort92a} \\
        & 40.8 & +5/-5     & C$_2$H$_6$ & \citet{ort90} \\
Uranus  & No data & \\
Neptune & 87   & +26/-26   & C$_2$H$_6$ & \citet{sad96} \\
        & 78   & +26/-26   & C$_2$H$_6$ & \citet{ort92b} \\
        & 74.1 & +23/-23   & C$_2$H$_6$ & \citet{ort90} 
\enddata
\end{deluxetable}

\subsection{Minor bodies of the Solar System}

See Tables~\ref{tab:minorbodies} and \ref{tab:comets} for data on
minor bodies and comets, respectively.

\begin{deluxetable}{lccl}
\tablecaption{\label{tab:minorbodies} $^{12}$C/$^{13}$C ratios in
  minor Solar System bodies.}  \tablehead{\colhead{Object} &
  \colhead{$^{12}$C/$^{13}$C} & \colhead{Error} &
  \colhead{Reference(s)}} 
\tablewidth{0pt}
\startdata 
Meteorites & 87.4 & +1.2/-1.2 & \citet{mar08} \\
           & 90.8 & +1.4/-1.4 & \citet{ale07} \\
           & 86.5 & +2.0/-2.0 & \citet{piz04} \\
           & 91   & +4/-4     & \citet{gra03} \\
           & 91.6 & +0.1/-0.1 & \citet{wri94} \\
           & 91.0 & +0.3/-0.3 & \citet{wri94} \\
           & 90.5 & +1.1/-1.1 & \citet{gra86} \\
           & 89   & --        & \citet{yan84} \\
           & 90.4 & +0.5/-0.5 & \citet{rob82} \\
           & 89   & +2/-2     & \citet{boa54} \\
Presolar grains & 3-5000 & --      & \citet{hop97} \\
IDPs       & 89   & +11/-11   & \citet{flo06} \\
           & 92   & +4/-4     & \citet{mes03} 
\enddata
\end{deluxetable}

\begin{deluxetable}{lcccl}
\tablecaption{\label{tab:comets} $^{12}$C/$^{13}$C ratios in comets.}
\tablehead{\colhead{Comet} & \colhead{$^{12}$C/$^{13}$C} &
  \colhead{Error} & \colhead{Molecule} & \colhead{Reference(s)}}
\tablewidth{0pt}
\tabletypesize{\footnotesize}
\startdata
81P/ Wild 2         & 92.2 & +1.4/-1.4 & C$^-$ & \citet{mck06}\\
C/17P/Holmes        & 114 & +26/-26 & HCN & \citet{boc08} \\
C/1996 B2 Hyakutake & 34 & +12/-12  & HCN & \citet{lis97b} \\
C/1995 O1 Hale Bopp & 94  & +8/-8   & HCN & \citet{boc08} \\
                    & 65  & +13/-13 & HCN & \citet{boc08} \\
                    & 109 & +22/-22 & HCN & \citet{ziu99} \\
                    & 111 & +12/-12 & HCN & \citet{jew97} \\
                    & 90\tablenotemark{a} & +30/-30 & CN  & \citet{man05} \\
                    & 95\tablenotemark{a} & +40/-40 & CN  & \citet{man05} \\
                    & 80\tablenotemark{a} & +20/-20 & CN  & \citet{man05} \\
                    & 90  & +20/-20 & CN  & \citet{man09,man05} \\
                    & 100 & +30/-30 & CN  & \citet{man09,man05} \\
                    & 80  & +25/-25 & CN  & \citet{man09,man05} \\
                    & 165 & +40/-40 & CN  & \citet{arp03} \\
                    & 90  & +15/-15 & CN  & \citet{lis97a} \\
1P/Halley           & 95  & +12/-12 & CN  & \citet{kle95} \\
                    & 89  & +17/-17 & CN  & \citet{jaw91} \\
                    & 65  & +9/-9   & CN  & \citet{wyc89} \\
C/2003 K4 LINEAR    & 90\tablenotemark{a}  & +15/-15 & CN  & \citet{man05} \\
                    & 85\tablenotemark{a}  & +20/-20 & CN  & \citet{man05} \\
                    & 90  & +20/-20 & CN  & \citet{man09,man05} \\
                    & 80  & +20/-20 & CN  & \citet{man09,man05} \\
C/1990 K1 Levy      & 90  & +10/-10 & CN  & \citet{wyc00} \\
C/1989 X1 Austin    & 85  & +20/-20 & CN  & \citet{wyc00} \\
C/1989 XIX Okazaki-Levy-Rudenko & 93 & +20/-20 & CN & \citet{wyc00} \\
C/2001 Q4 NEAT      & 90  & +15/-15 & CN  & \citet{man05} \\
C/2001 Q4 NEAT      & 70  & +30/-30 & CN  & \citet{man05} \\
C/2000 WM1 LINEAR   & 115\tablenotemark{a} & +20/-20 & CN  & \citet{arp03} \\
                    & 100 & +20/-20 & CN  & \citet{man09,arp03} \\
C/1999 S4 LINEAR    & 100\tablenotemark{a} & +30/-30 & CN  & \citet{hut05} \\
                    & 90  & +30/-30 & CN  & \citet{man09,hut05} \\
88P/1981 Q1 Howell  & 90  & +10/-10 & CN  & \citet{hut05} \\
122P/1995 S1 de Vico & 90  & +10/-10 & CN  & \citet{jeh04} \\
153P/2002 C1 Ikeya-Zhang & 90\tablenotemark{a} & +25/-25 & CN & \citet{jeh04} \\
                    & 80 & +30/-30 & CN & \citet{man09,jeh04} \\
9P/ Tempel 1        & 95 & +15/-15  & CN  & \citet{jeh06} \\
C/17P/Holmes        & 90 & +20/-20  & CN  & \citet{man09} \\
                    & 90\tablenotemark{a} & +20/-20  & CN  & \citet{boc08} \\
                    & 95 & +20/-20  & CN  & \citet{man09,boc08} \\
Kohoutek 1973 XII   & 115 & +30/-20 & C$_2$ & \citet{dan74} \\
                    & 135 & +60/-45 & C$_2$ & \citet{dan74} \\
Ikeya 1963 I        & 70  & +15/-15 & C$_2$ & \citet{sta64} \\
Tago-Sato-Kosaka    & 100 & +20/-20 & C$_2$ & \citet{van91} \\
Kobayashi-Berger-Milon & 110 & +20/-30 &  C$_2$ & \citet{van77} \\
West 1976 VI        & 60  & +15/-15 & C$_2$ & \citet{lam83} 
\enddata
\tablenotetext{a}{This value has been subsequently revised.}
\end{deluxetable}
\clearpage

\section{Statistics}

Here I mention some brief and simple statistics:
\begin{itemize}
\item The mean $^{12}$C/$^{13}$C ratio for the Sun is 85.0.
\item The mean $^{12}$C/$^{13}$C ratio for the rocky planets and the
  Moon is 89.8.
\item The mean $^{12}$C/$^{13}$C ratio for the gas-giant planets and
  Titan is 85.3.
\item The mean $^{12}$C/$^{13}$C ratio for comets is 93.4.
\item A weighted mean of all results is 91.0, which is influenced by
  the high number of cometary measurements with large error
  bars. \citet{sad96} found a similar value of 88.7 for the outer
  planets.
\end{itemize}

\acknowledgements

This research has made use of NASA's Astrophysics Data System
Bibliographic Services

\clearpage

\end{document}